\documentclass[%
 reprint,
 amsmath,amssymb,
 aps,
prb,
]{revtex4-2}

\usepackage[dvipdfmx]{graphicx}
\usepackage[dvipdfmx]{color}
\usepackage{dcolumn}
\usepackage{bm}
\usepackage{braket}

\newcommand{\QQ}{\mathcal{Q}}

\begin{document}

\title{Spin moir\'{e} engineering of topological magnetism and emergent electromagnetic fields
}

\author{Kotaro Shimizu, Shun Okumura, Yasuyuki Kato, and Yukitoshi Motome}
\affiliation{Department of Applied Physics, The University of Tokyo, Tokyo 113-8656, Japan}

\date{\today}

\begin{abstract}
A superposition of spin helices can yield topological spin textures, such as skyrmion and hedgehog lattices.
Based on the analogy with the moir\'e in optics, we study the magnetic and topological properties of such superpositions 
in a comprehensive way by modulating the interference pattern continuously. 
We find that the control of the angles between the superposed helices and 
the net magnetization yields successive topological transitions associated with pair annihilation 
of hedgehogs and antihedgehogs. 
Accordingly, emergent electromagnetic fields, magnetic monopoles and antimonopoles, and Dirac strings 
arising from the noncoplanar spin textures show systematic evolution.
In addition, we also show how the system undergoes the magnetic transitions with dimensional reduction 
from the three-dimensional hedgehog lattice to a two-dimensional skyrmion lattice or a one-dimensional conical state. 
The results indicate that the concept of ``spin moir\'{e}" provides an efficient way of engineering 
the emergent electromagnetism and topological nature in magnets. 
\end{abstract}


\maketitle

\section{Introduction \label{sec:1}}
A superposition of waves generates interference fringes called moir\'{e}. 
As the moir\'{e} patterns depend on, e.g., the periods, amplitudes, phases, and propagating angles 
of the superposed waves, manipulation of such parameters allows us to control the superstructures for many applications, 
especially in optical phenomena~\cite{Oster1964,Sciammarella1982}.
Recently, moir\'{e} patterns have attracted renewed interests in condensed matter physics, especially 
in atomic layer materials.  
There, the two-dimensional (2D) array of atoms causes interferences in stacking structures, and 
the resultant superlattices bring up new length and energy scales. 
The highlight is twisted graphenes, where the twist angle between different layers modulates 
the low-energy electronic state~\cite{Lopes2007,Trambly2010,Suarez2010,Kim3364} and 
drives emergent quantum phenomena such as superconductivity~\cite{Cao2018-1} and correlated insulators~\cite{Cao2018-2}.

Similar moir\'{e} patterns are expected in magnetism for superpositions of spin density waves, 
which are often called multiple-$Q$ spin states. 
The typical examples are the so-called skyrmion lattice (SkL)~\cite{Muhlbauer2009,Yu2010} 
and hedgehog lattice (HL)~\cite{Kanazawa2012,Tanigaki2015,Fujishiro2019,Ishiwata2020}. 
In many cases, these states are studied for particular choices of the superposed waves corresponding to 
each situation, but given the analogy with the moir\'e in optics, one can conceive a further variety of 
the spin patterns and their continuous modulations. 
Moreover, such ``spin moir\'{e}'' picture would bring about richer physics compared to the structural ones in atomic layer materials. 
First of all, the spin moir\'es have larger degrees of freedom since the constituent waves are vector fields, 
in contrast to the scalar ones for the density waves of atoms and charges.
Second, they are not restricted to two dimensions; actually, the HL is a three-dimensional (3D) moir\'{e}. 
Then, some of them are topologically protected, as represented by the skyrmions and hedgehogs. 
Furthermore, they can generate emergent electromagnetic fields through the Berry phase mechanism~\cite{Berry1984,Xiao2010}. 
Despite these virtues, most studies on these spin textures lack the viewpoint of the moir\'{e} physics. 
Once one can control the interferences like in optics, 
the spin moir\'{e}s would bring a new perspective on not only magnetism but also electronic, transport, and 
optical properties associated with the topological nature and the emergent electromagnetic fields.

In this paper, we theoretically explore the moir\'{e} physics in a superposition of multiple spin helices. 
Taking a 3D HL as an archetypal example, we study the systematic evolution of the magnetic and topological properties 
by tuning the moir\'{e} pattern through the angles between the helical directions and the net magnetization. 
Tracking the real-space positions of hedgehogs and antihedgehogs, we find that the moir\'{e} manipulation drives 
successive topological transitions caused by their pair annihilation. 
We find that the net emergent electric and magnetic fields show anomalies at the topological transitions.  
We also show that the emergent magnetic field, 
which is directly related with the topological Hall response,  
is given by the sum of the projections of Dirac strings connecting the magnetic monopoles and antimonopoles. 
Furthermore, assuming the Ginzburg-Landau free energy, we demonstrate that 
the increase of the net magnetization changes the helical angles, resulting in not only the topological transitions but also 
magnetic phase transitions with dimensional reduction from the 3D HL to a 2D SkL or a one-dimensional (1D) conical state. 
Our results would pave the way for engineering the emergent electromagnetism in magnetic materials 
from the viewpoint of the moir\'{e} physics. 

The rest of the paper is organized as follows. 
In Sec.~\ref{sec:2}, we introduce the set up of the 3D spin moir\'{e} and exemplify the spin patterns 
by varying the angle spanning the helical directions and the uniform magnetization. 
In Sec.~\ref{sec:3}, after introducing the hedgehogs and antihedgehogs, the Dirac strings, and 
the emergent electromagnetic fields (Sec.~\ref{sec:3.1}),
we discuss how these topological properties evolve through the spin moir\'{e} modulation (Sec.~\ref{sec:3.2}). 
In Sec.~\ref{sec:4}, we present the results of variational calculations based on the Ginzburg-Landau free
energy. 
We first describe the details of the variational calculations (Secs.~\ref{sec:4.1} and \ref{sec:4.2}), and then, 
we show two representative cases while increasing the magnetization (Sec.~\ref{sec:4.3}). 
We discuss the results in Sec.~\ref{sec:5}. Section~\ref{sec:6} is devoted to the summary of this paper.

\section{Spin moir\'e: set up \label{sec:2}}

\begin{figure}[t]
	\includegraphics[width=1.0\columnwidth]{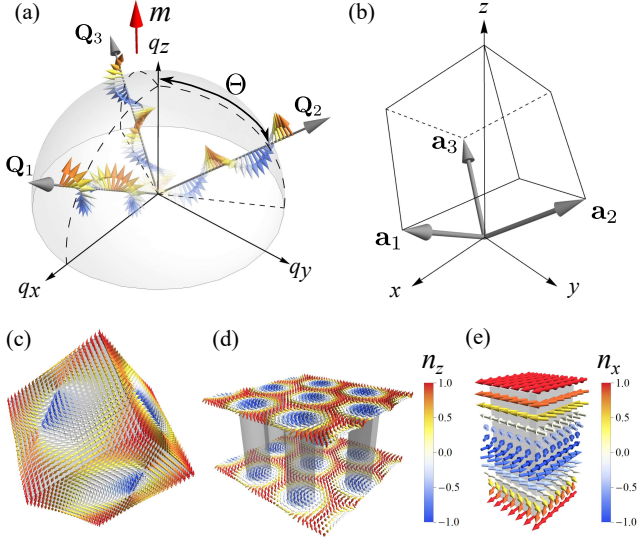}
	\caption{
	\label{fig:setup}
	(a) Schematic of the superposition of three helices in Eq.~(\ref{eq:3Qansatz}). 
	(b) Corresponding rhombohedral magnetic unit cell. 
	(c)-(e) Real-space spin configurations of (c) the 3D HL, (d) the 2D SkL, 
	(e) and the 1D helical state, which are obtained at $\Theta=\Theta_{\rm cubic}$, 
	$\pi/2$, and $0$ in Eq.~(\ref{eq:3Qansatz}) with $m=0$, respectively.
	The arrows represent the normalized spins ${\bf  n}({\bf r})={\bf S}({\bf r})/|{\bf S}({\bf r})|$, 
	whose color denotes the $z$ component in (c) and (d), and the $x$ component in (e).
	}
\end{figure} 

We consider 3D spin moir\'{e} patterns generated by a superposition of three spin helices in continuous space, 
focusing on the control by the angle spanning the helical directions and 
the magnetization induced by an external magnetic field. 
The set up is schematically shown in Fig.~\ref{fig:setup}(a).
Specifically, we assume the spin ${\bf S}({\bf r})$ at position ${\bf r}$ in continuous space 
in the form 
\begin{eqnarray}
	{\bf S}({\bf r}) =
	\sum_{\eta=1}^{3} \frac{1}{\sqrt{3}}\left(
	{\bf e}_{\eta}^{1}\cos\QQ_{\eta}
	+{\bf e}_{\eta}^{2}\sin\QQ_{\eta}
	\right)+m\hat{{\bf z}}, \label{eq:3Qansatz}
\end{eqnarray}
where 
$\QQ_{\eta}={\bf Q}_{\eta}\cdot{\bf r}$ and ${\bf Q}_{\eta}$ denote the three helical wave vectors as 
${\bf Q}_{\eta}=Q{\bf e}_{\eta}^{0}$~\footnote{
We ignore the phase degrees of freedom except for the 2D SkL with $\Theta=\pi/2$ in the variational calculations; see Sec.~\ref{sec:4B2}.
To make the 1D helical and conical state with $\Theta=0$, we need to choose proper phases, but this is irrelevant for the variational calculations. 
}. 
In this study, we take 
\begin{eqnarray}
{\bf e}_1^0&=&\left(\sin\Theta,0,\cos\Theta\right)
,  \\
{\bf e}_2^0&=&\left(-\frac{1}{2}\sin\Theta, \frac{\sqrt{3}}{2}\sin\Theta,\cos\Theta\right)
, \\ 
{\bf e}_3^0&=&\left(-\frac{1}{2}\sin\Theta,-\frac{\sqrt{3}}{2}\sin\Theta,\cos\Theta\right), 
\end{eqnarray}
and ${\bf e}_\eta^0$, ${\bf e}_\eta^1$, and ${\bf e}_\eta^2$ as the orthogonal unit vectors satisfying 
${\bf e}_{\eta}^2={\bf e}_{\eta}^0\times{\bf e}_{\eta}^1$ with 
\begin{eqnarray}
{\bf e}_1^1&=&\left(-\cos\Theta,0,\sin\Theta\right) \label{eq:e_1^1}
, \\
{\bf e}_2^1&=&\left(\frac{1}{2}\cos\Theta,-\frac{\sqrt{3}}{2}\cos\Theta,\sin\Theta\right) \label{eq:e_2^1}
, \\
{\bf e}_3^1&=&\left(\frac{1}{2}\cos\Theta,\frac{\sqrt{3}}{2}\cos\Theta,\sin\Theta\right). \label{eq:e_3^1}
\end{eqnarray}
Thus, the helical planes spanned by ${\bf e}_\eta^1$ and ${\bf e}_\eta^2$ are perpendicular to the helical directions ${\bf e}_\eta^0$, meaning that Eq.~(\ref{eq:3Qansatz}) is the superposition of the proper-screw helices  (see the end of this section and Appendix). 
The last term in Eq.~(\ref{eq:3Qansatz}) is the uniform magnetization along  
$\hat{\bf z} = (0,0,1)$.
The magnetic unit cell (MUC) of this spin texture for $0<\Theta<\pi/2$ is a rhombohedral one shown in Fig.~\ref{fig:setup}(b)
whose translation vectors ${\bf a}_{\eta}$ are defined by 
${\bf a}_{\eta}\cdot{\bf Q}_{\eta'}=2\pi\delta_{\eta,\eta'}$, where $\delta_{\eta,\eta'}$ is the Kronecker delta. 
The explicit forms are given as
\begin{eqnarray}
{\bf a}_1 &=& \frac{2\pi}{Q}\left( \frac{2}{3\sin\Theta}, 0 , \frac{1}{3\cos\Theta}\right), \\
{\bf a}_2 &=& \frac{2\pi}{Q}\left( -\frac{1}{3\sin\Theta}, \frac{1}{\sqrt{3}\sin\Theta}, \frac{1}{3\cos\Theta}\right), \\
{\bf a}_3 &=& \frac{2\pi}{Q}\left( -\frac{1}{3\sin\Theta}, -\frac{1}{\sqrt{3}\sin\Theta}, \frac{1}{3\cos\Theta} \right).
\end{eqnarray}
The volume of the MUC is given by 
\begin{equation}
V={\bf a}_1\cdot({\bf a}_2\times{\bf a}_3)=\left(\frac{2\pi}{Q}\right)^{3}
\frac{2}{3\sqrt{3}\sin^2\Theta \cos\Theta}.
\label{eq:V}
\end{equation}

We choose Eq.~(\ref{eq:3Qansatz}) so as to satisfy threefold rotational symmetry about the $z$ axis and 
to interpolate the three known spin textures while changing $\Theta$: 
(i) the 3D HL when ${\bf Q}_\eta$ are orthogonal to each other, namely, at 
\begin{equation}
\Theta=\Theta_{\rm cubic}=\arccos\left(\frac{1}{\sqrt{3}}\right),
\end{equation}
where the MUC becomes cubic, 
(ii) the 2D SkL at $\Theta=\pi/2$ where the three wave vectors ${\bf Q}_\eta$ lie on the $q_x q_y$ plane with  
$120^\circ$ relations, and (iii) 1D helical state ($m=0$) or conical state ($m>0$) at $\Theta=0$ (see also Sec.~\ref{sec:4}). 
The real-space spin structures of these three states are shown in 
Figs.~\ref{fig:setup}(c), \ref{fig:setup}(d), and \ref{fig:setup}(e).
Note that the 3D HL represents the one discovered in the $B$20 compound MnGe~\cite{Kanazawa2012,Tanigaki2015}, 
while the 2D SkL and the 1D helical or conical state are found in the other $B$20 compound MnSi~\cite{Muhlbauer2009,Yu2010}.

\begin{figure*}[t]
	\includegraphics[width=2.0\columnwidth]{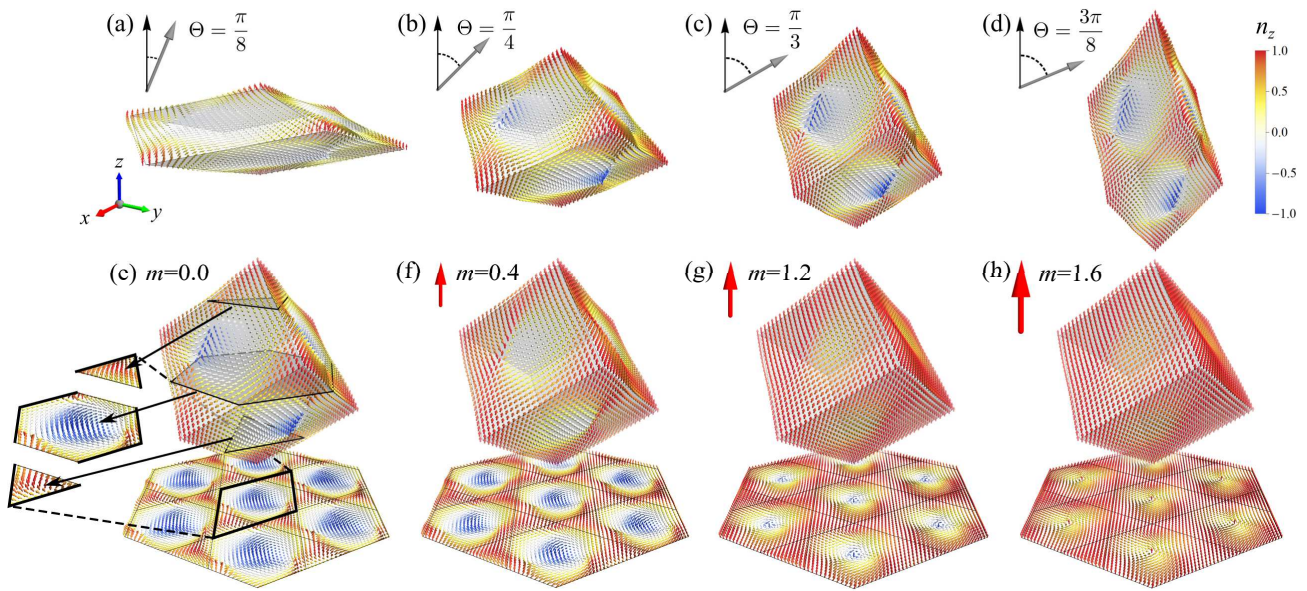}
	\caption{
		\label{fig:moirechange}
		Spin moir\'{e}s for (a) $\Theta=\frac{\pi}{8}$, (b) $\Theta=\frac{\pi}{4}$,  
		(c) $\Theta=\frac{\pi}{3}$, and (d) $\Theta=\frac{3\pi}{8}$ at $m=0$, and (e) $m=0$, (f) $m=0.4$,  
		(g) $m=1.2$, and (h) $m=1.6$ at $\Theta=\Theta_{\rm cubic}=\arccos(\frac{1}{\sqrt{3}})$. 
		The arrows represent the normalized spins ${\bf  n}({\bf r})={\bf S}({\bf r})/|{\bf S}({\bf r})|$, 
		whose color denotes the $z$ component.
		The lower panels in (e)-(h) show the horizontal slices through the center of the MUC. 
		The bold black rhombus in (e) represents the 2D MUC for the slices.
		}
\end{figure*}

The superposition in Eq.~(\ref{eq:3Qansatz}) allows a continuous modulation of the spin moir\'{e} patterns 
between these known spin textures by varying $\Theta$ and $m$.  
Figures~\ref{fig:moirechange}(a)-\ref{fig:moirechange}(d) exemplify such modulation while changing $\Theta$ at $m=0$. 
In this case, the MUC changes its shape and volume, together with the modulation of the spin moir\'{e}. 
Figures~\ref{fig:moirechange}(e)-\ref{fig:moirechange}(h) display the cases while changing $m$ at 
$\Theta=\Theta_{\rm cubic}$ where the MUC is cubic. 
We also present the spin patterns on the horizontal plane through the center of the MUC in the lower panels. 
The plane comprises a SkL; the skyrmion number is $-1$ per the 2D MUC 
shown by the black rhombus in Fig.~\ref{fig:moirechange}(e)~\cite{Nagaosa2013}.
While increasing $m$, the spin-up region is extended on the verge of skyrmions 
[Figs.~\ref{fig:moirechange}(e)-\ref{fig:moirechange}(g)], and eventually, 
the system turns into a topologically trivial state when the spins at the cores are flipped [Fig.~\ref{fig:moirechange}(h)]. 

While we take the proper screws in Eq.~(\ref{eq:3Qansatz}), we note that the tilting of the helical planes with respect to the helical directions is included in the following analysis as long as the threefold rotational symmetry about the $z$ axis is preserved. 
This is because the superposition of such tilted helices can be transformed into that of the proper screws with a different $\Theta$ in Eq.~(\ref{eq:3Qansatz}), as discussed in Appendix.

\section{Topological properties \label{sec:3}}

As exemplified in Fig.~\ref{fig:moirechange}, the angle $\Theta$ and the magnetization $m$ modulate 
the spin moir\'e pattern in Eq.~(\ref{eq:3Qansatz}). 
Accordingly, they modulate the topological nature, which is characterized by 
the hedgehogs and antihedgehogs, the Dirac strings, and the emergent electromagnetic fields. 
We discuss the evolution of the topological properties in this section.
We define the topological objects and the emergent electromagnetic fields in Sec.~\ref{sec:3.1}. 
In Sec.~\ref{sec:3.2}, we elucidate the topological phase diagram which shows 
how the topological properties evolve while changing $\Theta$ and $m$ in Eq.~(\ref{eq:3Qansatz}).

\subsection{Topological objects and emergent electromagnetic fields \label{sec:3.1}}

\subsubsection{Hedgehog and antihedgehog}

The spin structure in Eq.~(\ref{eq:3Qansatz}) for $0<\Theta<\pi/2$ is regarded as the HL composed of three helices ($3Q$-HL) 
as it has a periodic array of the topological defects, hedgehogs and antihedgehogs. 
In Eq.~(\ref{eq:3Qansatz}), the hedgehogs and antihedgehogs are defined as the singular points 
where the spin length vanishes, namely ${\bf S}({\bf r})=0$. 
They are regarded as the magnetic monopoles and antimonopoles, i.e., 
the sources and sinks of the emergent magnetic field, which is defined by
${\bf b}({\bf r}) = (b_x({\bf r}), b_y({\bf r}), b_z({\bf r}))$ with 
\begin{equation}
b_{i}({\bf r}) = \frac{1}{2}\varepsilon^{ijk} {\bf n}({\bf r})\cdot( \partial_{j} {\bf n}({\bf r}) \times \partial_{k} {\bf n}({\bf r}) ),
\label{eq:b_cont}
\end{equation}
where $\varepsilon^{ijk}$ is the Levi-Civita symbol and ${\bf n}({\bf r}) = {\bf S}({\bf r})/|{\bf S}({\bf r})|$ 
is the normalized spin~\cite{Volovik1987,Nagaosa2013}; here, the Einstein summation convention is used for repeated indices~\cite{EMconv}. 
The topological number called the monopole charge is defined by 
\begin{equation}
Q_{\rm m}=\frac{1}{4\pi}\int d\boldsymbol{\mathcal{S}} \cdot {\bf b}({\bf r}),
\label{eq:Qm_cont}
\end{equation} 
whose integral is taken on a closed surface surrounding the defect~\cite{Volovik1987,Kanazawa2016}; 
$Q_{\rm m}$ takes a positive (negative) integer for (anti)monopoles~\footnote{In the numerical calculations of 
$Q_{\rm m}$, we replace the integral in Eq.~(\ref{eq:Qm_cont}) by a discrete summation on eight corners of a small
cube which we define around the defect, and adopt the definition for a discrete lattice 
used in Ref.~\cite{Okumura2020}, which gives in principle the same value as Eq.~(\ref{eq:Qm_cont}) with Eq.~(\ref{eq:b_cont}).}.

\begin{figure*}[t]
\centering
\includegraphics[width=1.95\columnwidth]{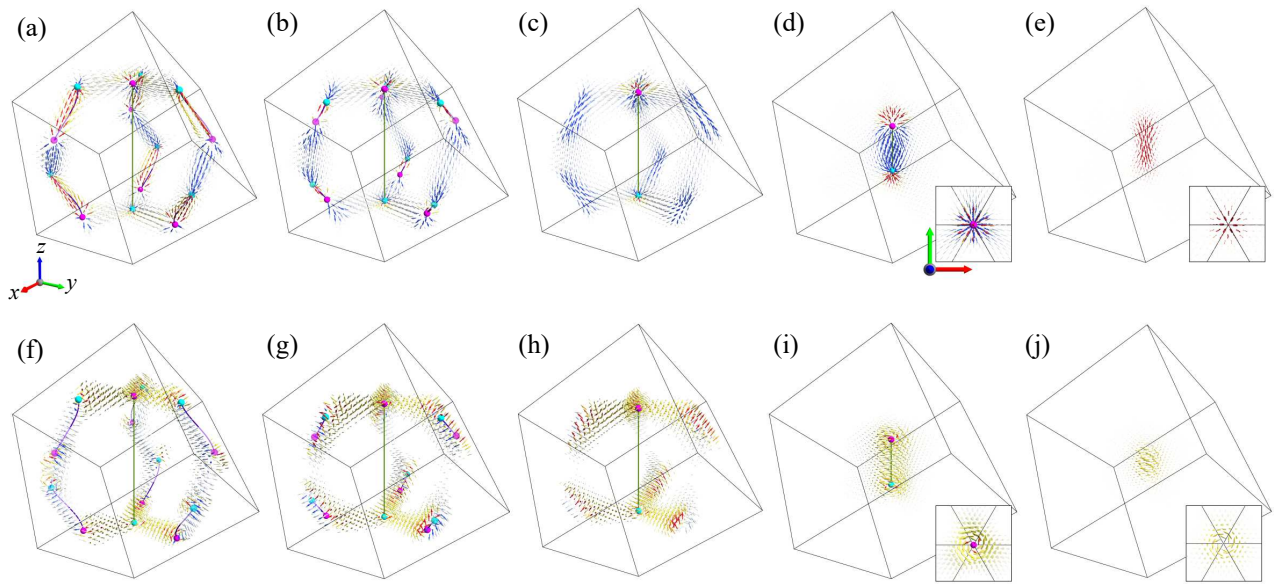}
\caption{ \label{fig:eemf_vec}
	Monopoles and antimonopoles, Dirac strings, and emergent electromagnetic fields, in the spin moir\'{e} 
	in Eq.~(\ref{eq:3Qansatz}) while changing $m$ at $\Theta=\Theta_{\rm cubic}$: 
	(a)(f) $m=0$, (b)(g) $m=0.24$, (c)(h) $m=0.4$, (d)(i) $m=1.2$, and (e)(j) $m=1.6$. 
	The arrows in (a)-(e) and (f)-(j) display the emergent magnetic and electric fields, 
	${\bf b}({\bf r})$ and ${\bf e}({\bf r})$, respectively. 
	Only top 15\% and 10\% in lengths are shown for ${\bf b}({\bf r})$ and ${\bf e}({\bf r})$, respectively, 
	and the divergent lengths close to the monopoles and antimonopoles are normalized appropriately for clarity. 
	The magenta and cyan spheres are the hedgehogs and antihedgehogs, and the green and purple  lines denote 
	the Dirac strings connecting them with the vorticity $\zeta=+1$ and $-1$, respectively. 
	The pale colors represent the objects in the neighboring MUCs. 
	The colors of the arrows are guides for the $z$ components. 
	The insets in (d), (e), (i), and (j) are the projections of the central part onto the $xy$ plane. 
}
\end{figure*}

In the absence of the magnetization $m=0$, the real-space positions of the hedgehogs and antihedgehogs 
are obtained analytically. 
Solving the equation ${\bf S}({\bf r})=0$ for Eq.~(\ref{eq:3Qansatz}), we obtain the following eight solutions: 
\begin{eqnarray}
\QQ_{\eta}^{*}
&=&
\left(\frac{\pi}{2},\ \frac{\pi}{2},\ \frac{\pi}{2} \right),\ \left(\frac{3\pi}{2},\ \frac{3\pi}{2},\ \frac{3\pi}{2} \right), \notag \\
&&\left(\frac{\pi}{2},\ \frac{\pi}{2}+2p_0(\Theta),\ \frac{\pi}{2}-2p_0(\Theta) \right)\notag \\ 
&&\qquad \qquad \qquad \mbox{and cyclic permutations}, \notag \\
&&\left(\frac{3\pi}{2},\ \frac{3\pi}{2}+2p_0(\Theta),\ \frac{3\pi}{2}-2p_0(\Theta) \right)\notag \\ 
&&\qquad \qquad \qquad \mbox{and cyclic permutations}, 
\label{eq:Q_eta^*}
\end{eqnarray}
where $\tan p_0(\Theta) = \sqrt{3}\cos\Theta$.
By using the relation 
\begin{eqnarray}
{\bf r}^{*} =(x^{*},y^{*},z^{*}) 
= \sum_{\eta} \frac{\QQ_{\eta}^{*}}{2\pi} {\bf a}_{\eta}, 
\end{eqnarray}
we obtain the real-space positions of the eight defects as
\begin{eqnarray}
{\bf r}^* &=&
\frac{2\pi}{Q}\left( 0, 0, \frac{1}{4\cos\Theta} \right),\ 
\frac{2\pi}{Q}\left( 0, 0, \frac{3}{4\cos\Theta} \right),\notag \\
&&
\frac{2\pi}{Q}\left( 0, \frac{2p_0(\Theta)}{\sqrt{3}\pi\sin\Theta}, \frac{1}{4\cos\Theta} \right)\notag \\ 
&&
\qquad\qquad\qquad\mbox{and $C_3^z$ symmetric points},\notag \\  
&&
\frac{2\pi}{Q}\left( 0, \frac{2p_0(\Theta)}{\sqrt{3}\pi\sin\Theta}, \frac{3}{4\cos\Theta} \right)\notag \\ 
&&
\qquad\qquad\qquad \mbox{and $C_3^z$ symmetric points}, 
\label{eq:defect_xyz_m=0}
\end{eqnarray}
where the $C_3^z$ symmetric points are obtained by $2\pi/3$ and $4\pi/3$ rotations around the $z$ axis.
The first two out of the eight solutions are located on the $z$ axis, 
while the rest six form a triangular prism around the $z$ axis 
[the locations will be depicted in Figs.~\ref{fig:eemf_vec}(a) and \ref{fig:eemf_vec}(f)]. 
As shown in Sec.~\ref{sec:3.2}, these defects are classified into four hedgehogs and four antihedgehogs 
according to the values of the monopole charge $Q_{\rm m}$ in Eq.~(\ref{eq:Qm_cont}).

For nonzero $m$, these defects change their positions. 
We cannot obtain all the solutions analytically, but find at least two as 
\begin{align}
&&
\QQ_\eta^* = 
\left(\frac{\pi}{2}+p(\Theta, m)
,\ \frac{\pi}{2}+p(\Theta, m)
,\ \frac{\pi}{2}+p(\Theta, m) 
\right), \notag \\
&&
\ \ \left(\frac{3\pi}{2}-p(\Theta, m)
,\ \frac{3\pi}{2}-p(\Theta, m)
,\ \frac{3\pi}{2}-p(\Theta, m) 
\right), 
\label{eq:Q_eta^*_m}
\end{align}
where $p(\Theta,m)=\arcsin\left(\frac{m}{\sqrt{3}\sin\Theta}\right)
$ $(|m|<\sqrt{3}\sin\Theta)$.
The real-space positions of these two defects are given by 
\begin{eqnarray}
{\bf r}^* = && 
\frac{2\pi}{Q}\left( 0, 0, \frac{1}{4\cos\Theta}\left(1 + \frac{2}{\pi}p(\Theta,m) \right) \right),\notag \\
&&
\frac{2\pi}{Q}\left( 0, 0, \frac{3}{4\cos\Theta}\left(1 - \frac{2}{\pi}p(\Theta,m) \right) \right).
\label{eq:defect_xyz_m}
\end{eqnarray} 
Note that these two defects corresponds to the first two in Eq.~(\ref{eq:defect_xyz_m=0}) on the $z$ axis for $m=0$. 
Equation~(\ref{eq:defect_xyz_m}) indicates that they move along the $z$ axis while increasing 
$m$ and pair annihilate at $|m|=\sqrt{3}\sin\Theta$. 
We obtain the other six solutions corresponding to the latter six in Eq.~(\ref{eq:defect_xyz_m=0}) numerically. 
As in the case with $m=0$, the eight defects are classified into four hedgehogs and four antihedgehogs. 
We confirm that any other solutions do not appear in the following calculations.

\subsubsection{Dirac string}

Next, we introduce the Dirac strings, which are the lines connecting hedgehogs and antihedgehogs.
Let us introduce the vector potential ${\bf a}({\bf r})$ for the emergent magnetic field ${\bf b}({\bf r})$ as 
${\bf b} ({\bf r}) = \bm{\nabla}\times{\bf a}({\bf r})$.
${\bf a}({\bf r})$ is gauge dependent. 
We choose the representation 
\begin{eqnarray}
	{\bf a}({\bf r}) = \left\{
	\begin{array}{l}
		{\bf a}^{\rm S}({\bf r}) = 
		\bigl(1-\cos\theta({\bf r})\bigr)\bm{\nabla} \phi({\bf r})\\
		{\bf a}^{\rm N}({\bf r}) = 
		-\bigl(1+\cos\theta({\bf r})\bigr)\bm{\nabla} \phi({\bf r}),
	\end{array}
	\right.
\label{eq:a}
\end{eqnarray} 
where $\theta({\bf r})$ and $\phi({\bf r})$ are the angular coordinates of the normalized spin as 
${\bf n}({\bf r})=
\left(\sin\theta({\bf r})\cos\phi({\bf r}), \sin\theta({\bf r})\sin\phi({\bf r}), \cos\theta({\bf r}) \right)$. 
Note that these two vector potentials are related with each other by the gauge transformation 
${\bf a}^{\rm N}({\bf r}) = {\bf a}^{\rm S}({\bf r}) - 2\bm{\nabla}\phi({\bf r})$.
In Eq.~(\ref{eq:a}), ${\bf a}^{\rm S}({\bf r})$ and ${\bf a}^{\rm N}({\bf r})$ have singularities at 
$\theta({\bf r})=\pi$ and $0$, respectively. 
To define ${\bf b}({\bf r})$ properly, we avoid these singularities by analytic continuation with 
using ${\bf a}^{\rm S}({\bf r})$ for $0 \leq \theta({\bf r}) \leq \pi-\delta$ and 
${\bf a}^{\rm N}({\bf r})$ otherwise, where $\delta$ is an infinitesimal~\footnote{
$\delta$ is introduced just to define the gauge connection. It does not appear in the numerical calculations of the monopole charge and the velocity of the Dirac string.}. 

\begin{figure}[t]
\includegraphics[width=1.0\columnwidth]{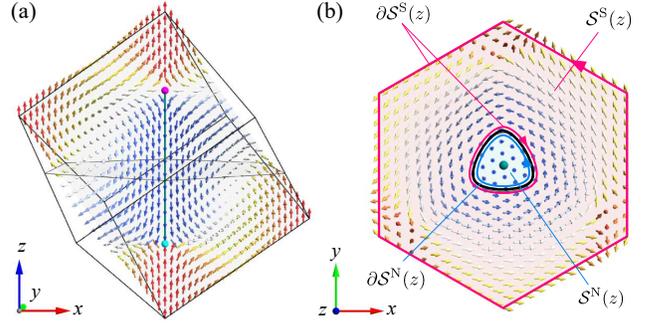}
\caption{
\label{fig:Diracstring}
(a) Spin configuration on the $xz$ slice through the center of the MUC for Eq.~(\ref{eq:3Qansatz}) 
with $\Theta=\Theta_{\rm cubic}$ and $m=0$.
The Dirac string with the vorticity $\zeta=+1$ (green vertical line) and the associated monopole and antimonopole 
(magenta and cyan spheres) are also shown.  
Note that only a single monopole-antimonopole pair out of four is shown for simplicity. 
(b) Spin configuration on the $xy$ slice through the center of the MUC. 
The green sphere at the center denotes the intersection of the Dirac string. 
The pink and blue areas denote $\mathcal{S}^{\rm S}(z)$ and $\mathcal{S}^{\rm N}(z)$ comprising 
$\mathcal{S}(z)$ in Eq.~(\ref{eq:bzave1}).
We take $\delta=\arccos(0.9)$ for better visibility.
The closed paths encircling the areas represent $\partial\mathcal{S}^{\rm S}(z)$ and $\partial\mathcal{S}^{\rm N}(z)$ 
in the first line of Eq.~(\ref{eq:bzave2}). 
}
\end{figure}

The singular points of the vector potentials form lines in real space. 
In the present system with the magnetization parallel to $\hat{\bf z}$, the singularity of ${\bf a}^{\rm S}({\bf r})$ 
where the spins point downward defines the Dirac strings connecting the monopoles and antimonopoles. 
The situation is exemplified for $\Theta=\Theta_{\rm cubic}$ 
and $m=0$ in Fig.~\ref{fig:Diracstring}, by taking the pair of monopole and antimonopole on the $z$ axis 
given by the first two solutions in Eq.~(\ref{eq:defect_xyz_m=0}). 
Figure~\ref{fig:Diracstring}(a) shows the $xz$ slice of the spin configuration through the center of the MUC; the Dirac string is denoted by the green vertical line connecting the monopole and antimonopole denoted by the magenta and cyan spheres, respectively. 
Figure~\ref{fig:Diracstring}(b) shows the $xy$ slice through the center of the MUC. 

The Dirac strings are distinguished by their vorticity given by 
\begin{equation}
\zeta = \frac{1}{2\pi} \oint d{\bf l} \cdot \bm{\nabla} \phi({\bf r}), 
\label{eq:vorticity}
\end{equation}
where the integral is taken on a closed path surrounding the string at ${\bf r}$~\footnote{In the 
numerical calculations of $\zeta$, we define a small square surrounding the Dirac string and count how many times the projections of the four spins on the corners onto the $xy$ plane wrap a unit circle.}. 
For the example in Fig.~\ref{fig:Diracstring}, the Dirac string has $\zeta=+1$.  
The value of $\zeta$ is, however, known without the direct calculation of Eq.~(\ref{eq:vorticity}), 
once we identify the monopole and antimonopole connected by the Dirac string; it is given by their relative positions 
and monopole charges $Q_{\rm m}$. 
For instance, for the Dirac string in Fig.~\ref{fig:Diracstring}, the monopole locates above the antimonopole in the $z$ direction and $Q_{\rm m} = \pm 1$, from which we can immediately assign $\zeta=+1$. 
We confirm that the numerical estimates are consistent with such topological assignments for all the Dirac strings in our analysis.  

\subsubsection{Emergent magnetic field}

The emergent magnetic field ${\bf b}({\bf r})$ is defined by a local noncoplanar spin configuration as in Eq.~(\ref{eq:b_cont}). 
For the bulk properties of the system, however, the net value of ${\bf b}({\bf r})$ in the MUC is important. 
For instance, the net emergent magnetic field leads to the topological Hall effect~\cite{Loss1992,Ye1999}. 
In the present case with the magnetization along the $z$ axis, as the $x$ and $y$ components of ${\bf b}({\bf r})$ 
cancel out due to threefold rotational symmetry about the $z$ axis, we introduce the spatially averaged value 
of the $z$ component as 
\begin{eqnarray}
\bar{b}_z = \frac{1}{4\pi L_z}\int d\mathcal{V}{b}_z({\bf r}), \label{eq:bzave}
\end{eqnarray}
where $L_z = | {\bf a}_1 + {\bf a}_2 + {\bf a}_3 |$ and the integral is taken within the MUC;
see Fig.~\ref{fig:setup}(b)~\cite{EMconv}. 
Note that $\bar{b}_z$ has the dimension of the magnetic flux.
Using the vector potential in Eq.~(\ref{eq:a}), we can write Eq.~(\ref{eq:bzave}) as 
\begin{equation}
\bar{b}_z = \frac{1}{4\pi L_z} \int dz \int_{\mathcal{S}(z)} d\mathcal{S} 
\hat{\bf z}\cdot\bigl(\bm{\nabla}\times{\bf a}({\bf r})\bigl),   
\label{eq:bzave1}
\end{equation}
whose integral is taken for a 2D horizontal slice of the MUC, $\mathcal{S}(z)$. 
Note that $\mathcal{S}(z)$ is divided into $\mathcal{S}^{\rm S}(z)$ and $\mathcal{S}^{\rm N}(z)$ 
corresponding to the regions where ${\bf a}({\bf r})$ is defined as 
${\bf a}^{\rm S}({\bf r})$ and ${\bf a}^{\rm N}({\bf r})$, respectively; see Fig.~\ref{fig:Diracstring}(b). 
Using Stokes' theorem under the periodic boundary conditions and Eq.~(\ref{eq:vorticity}), we obtain
\begin{eqnarray}
	\bar{b}_z 
	&=& \frac{1}{4\pi L_z}\int dz 
	\left[ 
		\oint_{\partial\mathcal{S}^{\rm S}(z)} d{\bf l}\cdot{\bf a}^{\rm S}({\bf r}) 
		+\oint_{\partial\mathcal{S}^{\rm N}(z)} d{\bf l}\cdot{\bf a}^{\rm N}({\bf r})
	\right] 
\notag \\
	&=& -\frac{1}{L_z} \sum_{k} \int dz \zeta_{k}
, \label{eq:bzave2}
\end{eqnarray}
where $k$ labels the Dirac strings and $\zeta_k
$ is the vorticity of the $k$th Dirac string; the last integral is taken in the $z$ range where $\zeta_k$ is nonzero.

In the present system, the vorticity takes $\zeta_k 
= +1$ or $-1$ for all the Dirac strings, as discussed later.
Thus, we can rewrite Eq.~(\ref{eq:bzave2}) into
\begin{equation}
	\bar{b}_z=-\frac{1}{L_z}\sum_k {\bf R}_k^{\pm}\cdot\hat{\bf z}, \label{eq:bzave3}
\end{equation}
where ${\bf R}^{\pm}_k$ denotes the vector directed from the antimonopole to the monopole 
at the ends of the $k$th Dirac string.
Equation~(\ref{eq:bzave3}) shows that $\bar{b}_z$ is given by the sum of the lengths of the Dirac strings 
projected onto the $z$ axis.

\subsubsection{Emergent electric field}

In addition, we introduce the emergent electric field by considering time ($t$) dependence in $m$ and $\Theta$ 
as ${\bf e}^m({\bf r}) \partial_t m + {\bf e}^\Theta({\bf r}) \partial_t \Theta$, whose $i=x,y,z$ component is defined by
\begin{equation}
e_{i}^{\nu}({\bf r}) 
= {\bf n}({\bf r})\cdot
( \partial_{i} {\bf n}({\bf r}) \times \partial_{\nu} {\bf n}({\bf r}) ),
\label{eq:ezave}
\end{equation}
where $\nu=m$ or $\Theta$~\cite{Volovik1987,Nagaosa2013}.
We also calculate the average of the $z$ component as 
\begin{eqnarray}
\bar{e}_z = \frac{L_z}{4\pi V}\int d\mathcal{V}\left[{e}_z^{m}({\bf r}) + {e}_z^\Theta({\bf r}) \partial_m \Theta\right] , \label{eq:ezave2}
\end{eqnarray}
where $V$ is the volume of the MUC in Eq.~(\ref{eq:V}) and the integral is taken within the MUC~\cite{EMconv}.
Note that $\bar{e}_z$ has the dimension of the electric voltage.

\subsection{Topological phase diagram \label{sec:3.2}}

\begin{figure}[t]
\includegraphics[width=1.0\columnwidth]{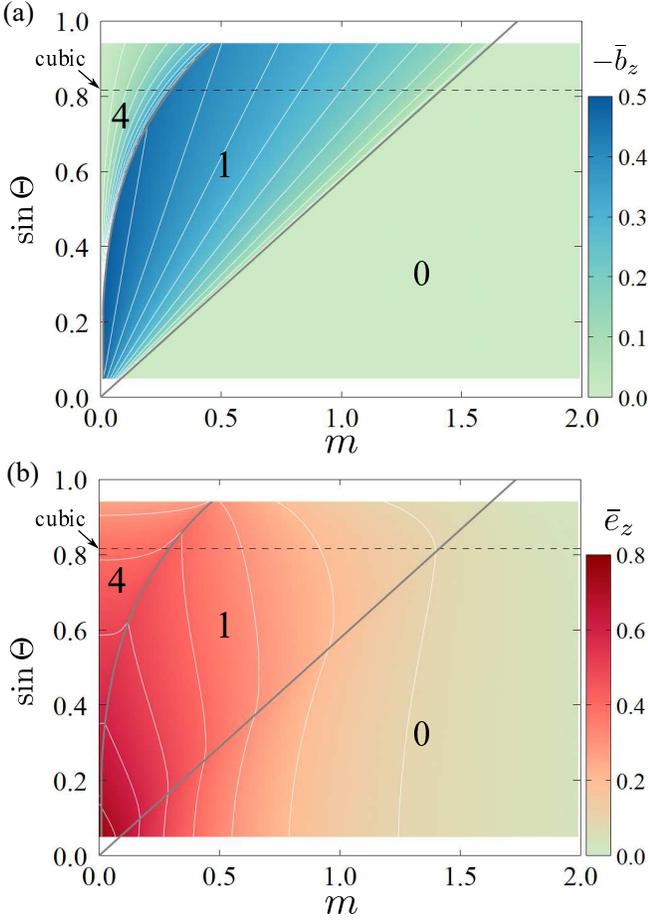}%
\caption{
\label{fig:eemf_cnt}
	Topological phase diagram determined by the number of monopole-antimonopole pairs within the MUC, $N_{\rm pair}$, 
	with the contour plots of (a) $-\bar{b}_z$ and (b) $\bar{e}_z$ as functions of $m$ and $\sin\Theta$. 
	$N_{\rm pair}$ is indicated in each phase divided by the gray boundaries. 
	The white lines denote the contours drawn every (a) $0.05$ and (b) $0.1$. 
	The blanks near $\sin\Theta=0$ and $1$ indicate the regions where sufficient numerical precision is not guaranteed. 
}
\end{figure}

Using the definitions in the previous subsection, we can identify and track the topological properties of the spin moir\'e in Eq.~(\ref{eq:3Qansatz}) while changing $\Theta$ and $m$. 
Figure~\ref{fig:eemf_vec} illustrates the systematic evolution of the topological objects and the emergent electromagnetic fields 
while increasing $m$ at $ \Theta=\Theta_{\rm cubic}$ in Eq.~(\ref{eq:3Qansatz}) 
[see the corresponding spin configurations in Figs.~\ref{fig:moirechange}(e)-\ref{fig:moirechange}(h)]. 
For $m=0$, the system has eight topological defects, as discussed in Eq.~(\ref{eq:Q_eta^*}). 
By computing the monopole charge $Q_{\rm m}$ in Eq.~(\ref{eq:Qm_cont}), we identify four out of them 
as monopoles with $Q_{\rm m}=+1$ and the rest four as antimonopoles with $Q_{\rm m}=-1$: 
The upper (lower) defect on the $z$ axis and the three on the lower (upper) plane of the triangular prism are (anti)monopoles. 
They are connected by the Dirac strings, as shown in Fig.~\ref{fig:eemf_vec}(a). 
Note that one of the Dirac strings runs straight along the $z$ axis, while the rest three are not straight and run through 
the boundaries of the MUC to connect with the (anti)monopoles in the neighboring MUCs; 
the former one has the vorticity $\zeta=+1$ and the latter three have $\zeta=-1$. 
The emergent magnetic field ${\bf b}({\bf r})$ in Eq.~(\ref{eq:b_cont}) flows from monopoles to antimonopoles. 
While increasing $m$, the monopoles and antimonopoles move toward their counterparts along the Dirac strings, 
as exemplified in 
Fig.~\ref{fig:eemf_vec}(b). 
With a further increase of $m$, three pairs connected by the Dirac strings with $\zeta=-1$ disappear with 
pair annihilation, leaving one pair connected by the Dirac string with $\zeta=+1$, as shown in Fig.~\ref{fig:eemf_vec}(c).
Finally, the remaining pair annihilates, leaving remnants of ${\bf b}({\bf r})$, as shown in 
Figs.~\ref{fig:eemf_vec}(d) and \ref{fig:eemf_vec}(e). 
These pair annihilations define topological transitions between different topological phases with different numbers of the monopole-antimonopole pairs (see Fig.~\ref{fig:eemf_cnt}). 

The corresponding results for the emergent electric field ${\bf e}({\bf r})$ in Eq.~(\ref{eq:ezave}) are shown in 
Figs.~\ref{fig:eemf_vec}(f)-\ref{fig:eemf_vec}(j).
Note that ${\bf e}^\Theta({\bf r})=0$ as $\Theta$ is fixed here.
As ${\bf e}({\bf r})$ and ${\bf b}({\bf r})$ follow  Faraday's law~\cite{Tatara2019}, 
${\bf e}({\bf r})$ becomes large where ${\bf b}({\bf r})$ changes rapidly while increasing $m$.
As a consequence, ${\bf e}({\bf r})$ appears along the flows of ${\bf b}({\bf r})$ with a swirling texture around them.

Figure \ref{fig:eemf_cnt} summarizes the topological phase diagram determined by 
the number of monopole-antimonopole pairs within the MUC, $N_{\rm pair}$, on the plane of $m$ and $\sin\Theta$. 
We find three different phases with $N_{\rm pair}=4$, $1$, and $0$. 
We also plot the averages of the emergent magnetic and electric fields, $\bar{b}_z$ in Eq.~(\ref{eq:bzave}) and 
$\bar{e}_z$ in Eq.~(\ref{eq:ezave2}), in Figs.~\ref{fig:eemf_cnt}(a) and \ref{fig:eemf_cnt}(b), respectively. 
Note that we plot $-\bar{b}_z$ in Fig.~\ref{fig:eemf_cnt}(a) and the contribution from $e_z^m({\bf r})$ in 
Fig.~\ref{fig:eemf_cnt}(b) for simplicity [we include $e_z^\Theta({\bf r})$ in the variational calculations in Sec.~\ref{sec:4}]. 
As shown in Fig.~\ref{fig:eemf_cnt}(a), $-\bar{b}_z$ rapidly increases with $m$ in the $N_{\rm pair}=4$ region, 
while decreases in the $N_{\rm pair}=1$ region after showing a sharp cusp on the boundary, 
and eventually vanishes in the $N_{\rm pair}=0$ region. 
In the $N_{\rm pair}=1$ region, using the relation in Eq.~(\ref{eq:bzave3}) and the real-space positions of 
the remaining pair in Eq.~(\ref{eq:defect_xyz_m}), 
$\bar{b}_z$ is explicitly obtained as a function of $m$ and $\Theta$ as 
\begin{eqnarray}
\bar{b}_z = -\frac12 \left[1-\frac{2}{\pi} \arcsin \left(\frac{m}{\sqrt{3}\sin\Theta}\right)\right].
\label{eq:bzave_Npair1}
\end{eqnarray}
Hence, the phase boundary from $N_{\rm pair}=1$ to $0$ 
is given by $m=\sqrt{3}\sin\Theta$. 
See also Eq.~(\ref{eq:defect_xyz_m}) and the discussion below it.
On the other hand, as shown in Fig.~\ref{fig:eemf_cnt}(b), $\bar{e}_z$ also shows a cusp at the boundary 
between the $N_{\rm pair}=4$ and $1$ regions, but it is large near the origin and 
decreases while increasing $m$ and $\Theta$. 
Note that $\bar{e}_z$ remains nonzero even in the topologically trivial state with $N_{\rm pair}=0$, in contrast to $\bar{b}_z$.

\section{Variational calculation \label{sec:4}}

In order to discuss how the system undergoes the topological transitions in Fig.~\ref{fig:eemf_cnt}, 
we need the energy functional in terms of $m$ and $\Theta$. 
Here, following the previous studies~\cite{Binz2006-1,Binz2006-2}, we assume the Ginzburg-Landau free energy 
to quartic order in $m$, and optimize it with respect to $\Theta$ while changing $m$ 
as a mimic of applying an external magnetic field. 

In this section, we first introduce the energy functional used in this study in Sec.~\ref{sec:4.1}. 
Then, we give the explicit forms of the energy for different spin states described by Eq.~(\ref{eq:3Qansatz}) in Sec.~\ref{sec:4.2}. 
Finally, we show the results of the variational calculations which show the evolution of 
$\Theta$, $N_{\rm pair}$, $\bar{b}_z$, and $\bar{e}_z$ while changing $m$ in Sec.~\ref{sec:4.3}.

\subsection{Energy functional \label{sec:4.1}}
In the present variational study, we introduce the Ginzburg-Landau free energy, following Refs.~\cite{Binz2006-1,Binz2006-2}. 
Specifically, we consider the energy functional given by $F=F_2+F_4$ with 
\begin{align}
F_2&=\sum_{{\bf q}} \left[ (r_0+Jq^2)|{\bf S}_{\bf q}|^2 +
2iD{\bf q}\cdot( {\bf S}_{\bf q} \times{\bf S}_{-{\bf q}} ) \right], 
\label{eq:F2}
\\
F_4&=\sum_{ {\bf q}_1 ,{\bf q}_2 ,{\bf q}_3 ,{\bf q}_4 } U({\bf q}_1,{\bf q}_2,{\bf q}_3) 
( {\bf S}_{{\bf q}_1}\cdot{\bf S}_{{\bf q}_2} )( {\bf S}_{{\bf q}_3}\cdot{\bf S}_{{\bf q}_4} )\notag \\
&\qquad\qquad\qquad \times \delta_{{\bf q}_1+{\bf q}_2+{\bf q}_3+{\bf q}_4, 0},
\label{eq:F4}
\end{align}
where ${\bf S}_{\bf q} = \frac{1}{V} \int d^3{\bf r} {\bf S}({\bf r}) e^{-i{\bf q}\cdot{\bf r}}$; 
$V$ is the volume of the MUC in Eq.~(\ref{eq:V}) and the integral is taken within the MUC. 
$F_2$ denotes the second-order interactions in terms of ${\bf S}_{\bf q}$: $r_0$, $J$, and $D$ represent 
the coefficients of the ${\bf q}$-independent bilinear, Heisenberg, and Dzyaloshinskii-Moriya interactions, 
respectively, and $q=|{\bf q}|$.
In the present calculation, we take $J>0$ and $D>0$; the Dzyaloshinskii-Moriya vector 
is assumed to be parallel to ${\bf q}$ so that it prefers a proper-screw spin structure.
$F_4$ represents the fourth-order interaction with the coupling constant $U({\bf q}_1,{\bf q}_2,{\bf q}_3)$; 
$\delta_{{\bf q}, {\bf q}'}$ is the Kronecker delta to impose the momentum conservation.
Assuming that $F_2$ is the leading term with $r_0-J\tilde{q}^2<0$ where $q=D/J \equiv \tilde{q}$, 
the system prefers a spin helix with the wave number $\tilde{q}$. 
The subleading $F_4$ tends to stabilize a noncollinear or noncoplanar spin texture composed of a superposition of the spin helices. 
In the following calculations, we take into account the spin structures represented by superpositions of 
the  helices with $\tilde{q} $ and the uniform ${\bf q}=0$ component; $\tilde{q} $ will be set at 
$|{\bf Q}_\eta|$ in the following calculations for the spin moir\'e in Eq.~(\ref{eq:3Qansatz}). 

Following Refs.~\cite{Binz2006-1,Binz2006-2}, we parametrize $U({\bf q}_1,{\bf q}_2,{\bf q}_3)$ with $|{\bf q}|=\tilde{q}$ 
by two parameters 
$\alpha=\frac{1}{2}\arccos( \hat{{\bf q}}_1\cdot\hat{{\bf q}}_2 )$ and  
$\beta=2\arccos\left[\frac{(\hat{{\bf q}}_2-\hat{{\bf q}}_1)\cdot\hat{{\bf q}}_3}{1-\hat{{\bf q}}_1\cdot\hat{{\bf q}}_2}\right]$, 
where $\hat{{\bf q}}={\bf q}/|{\bf q}|$, and assume the form of  
\begin{align}
U(\alpha,\beta)&=U_0+U_{11}\sin\alpha\cos\beta + U_{20}(3\cos^2\alpha-1) \notag \\
	&\quad+U_{22}\sin^2\alpha\cos2\beta.
\label{eq:U_alpha_beta}
\end{align}
In Eq.~(\ref{eq:U_alpha_beta}), the first term $U_0$ includes the contributions from several types of quartic interactions,  
but here, we assume that it is dominated by $W\bigl({\bf S}({\bf r})\cdot{\bf S}({\bf r})\bigl)^2$, for simplicity; namely we take $U_0=W$. 
In the following calculations, we leave $U_{11}$ and $U_{22}$ as parameters, and set $U_{20}=0$.
In addition, we take into account only the components $U(0,0,0)$, $U({\bf q},-{\bf q},0)$, and $U({\bf q},0,0)$ in $F_4$, 
and adopt the approximation 
$U(0,0,0) \approx U({\bf q}, -{\bf q}, 0)$.
Noting that $U({\bf q}, -{\bf q}, 0) \approx U(\alpha=\pi/2,\beta=0)$, 
we introduce the parameter $U_s$ which will be used in the next subsection as 
\begin{equation}
U\left(\alpha=\frac{\pi}{2},\beta=0\right) = U_0+U_{11}+U_{22} \equiv U_s. 
\end{equation}
All the simplifications and assumptions made above follow Refs.~\cite{Binz2006-1,Binz2006-2}. 

Then, by taking $W$ as the energy unit, the energy functional has three independent parameters, 
$U_{11}$, $U_{22}$, and $U({\bf q},0,0)$, which are denoted as 
\begin{equation}
	t=\frac{U_{11}}{W}, \ \
	u=\frac{U_{22}}{W}, \ \ 
	v=\frac{U({\bf q},0,0)}{W} . 
\label{eq:tuv}
\end{equation}
The parameters $r_0$, $J$, and $D$ in Eq.~(\ref{eq:F2}) are irrelevant, since the corresponding energy 
does not change for the different spin textures considered here within the variational state in Eq.~(\ref{eq:3Qansatz}), as shown below. 

\subsection{Energy for each state \label{sec:4.2}}

Using the functional introduced above, we estimate the energy for different spin states expressed by Eq.~(\ref{eq:3Qansatz}): 
the 3D HL for $0<\Theta<\pi/2$, the 2D SkL at $\Theta=\pi/2$, and the 1D helical or conical state at $\Theta=0$. 
We note that Eq.~(\ref{eq:3Qansatz}) cannot represent a uniformly polarized state which is expected to compete 
with Eq.~(\ref{eq:3Qansatz}) in the large $m$ region. 
For the energy comparison, we need specific values of $J$ and $D$ in Eq.~(\ref{eq:F2}), which are irrelevant to 
the other three spin states. 
Because of this complexity, we omit the uniformly polarized state in the following analysis.

\subsubsection{Hedgehog lattice}

First, we calculate the energy for the 3D HL given by Eq.~(\ref{eq:3Qansatz}) with $0<\Theta<\frac{\pi}{2}$. 
The energy is obtained as 
\begin{eqnarray} 
F_{\rm HL} 
&=& \sum_{\eta} [ f_2({\bf Q}_{\eta}) + f_s({\bf Q}_{\eta})] 
+ \sum_{\eta<\eta'} f_p( {\bf Q}_{\eta}, {\bf Q}_{\eta'}) \notag \\
&&
+ f_{m}^{(1)}(m)
+ \sum_{\eta} f_{m}^{(2)}( {\bf Q}_{\eta}, m ),
\label{eq:F_HL}
\end{eqnarray} 
where 
\begin{eqnarray} 
	&&f_2({\bf Q}_{\eta}) = \left( r_0 - J\tilde{q}^2 \right)|\psi_{{\bf Q}_{\eta}}|^2, \\
	&&f_s({\bf Q}_{\eta}) = U_s|\psi_{{\bf Q}_{\eta}}|^4, \\
	&&f_p( {\bf Q}_{\eta}, {\bf Q}_{\eta'} ) = 
	2V_p\left( \alpha_{\eta\eta'} \right)
	|\psi_{{\bf Q}_{\eta}}|^2|\psi_{{\bf Q}_{\eta'}}|^2, \\ 
	&&f_m^{(1)}(m) = r_0m^2+U_sm^4, \\
	&&f_m^{(2)}( {\bf Q}_{\eta}, m) = 
	2\left[U_sm^2 + U({\bf Q}_{\eta},0,0){\bf m}_{\perp,\eta}^2 \right]|\psi_{{\bf Q}_{\eta}}|^2, 
	\notag \\
	\label{eq:f^2_m}
\end{eqnarray}
with
\begin{eqnarray}
	&&\psi_{{\bf Q}_{\eta}}=\sqrt{2}|{\bf S}_{{\bf Q}_{\eta}}|, \\
	&&V_p(\alpha) = U \left( \frac{\pi}{2}, 4\alpha\right) +  U(\alpha,0)\sin^4\alpha  \notag \\
	&& \qquad\qquad + U \left( \frac{\pi}{2}-\alpha,0 \right)\cos^4\alpha, \\ 
	&&
	\alpha_{\eta\eta'} = {\frac12}\arccos \left( \hat{{\bf Q}}_{\eta} \cdot \hat{{\bf Q}}_{\eta'} \right), \\
	&&{\bf m}_{\perp,\eta}=m\left[\hat{\bf z}-\left(\hat{\bf z}\cdot\hat{{\bf Q}}_{\eta}\right)\hat{{\bf Q}}_{\eta}\right]. 
\end{eqnarray}
In this case, the following relations hold: 
\begin{equation}
|\psi_{{\bf Q}_{\eta}}|=\frac{1}{\sqrt{3}},\ \ 
\sin\alpha_{\eta\eta'} = \frac{\sqrt{3}}{2} \sin\Theta,
\label{eq:HL_conditions}
\end{equation}
where $0<\alpha_{\eta\eta'}<\pi/3$. 
In this energy functional, $v$ in Eq.~(\ref{eq:tuv}) couples $m$ and $\Theta$; Eq.~(\ref{eq:f^2_m}) indicates that 
a positive (negative) $v$ favors (disfavors) helices propagating in the $z$ direction, 
which tends to decreases (increases) $\Theta$ as $m$ increases. 

We note that the HL turns into a topologically trivial state with $N_{\rm pair}=0$ for 
$m>\sqrt{3}\sin\Theta$, as shown in Sec.~\ref{sec:3.2}, while the energy can be estimated 
by the same equation in Eq.~(\ref{eq:F_HL}); see also the results in Sec.~\ref{sec:4.3}.
We call such a topologically trivial state with a 3D spin texture the 3D $3Q$ state in the following analysis 
to distinguish it from the 2D  one derived from the SkL below.

\subsubsection{Skyrmion lattice}
\label{sec:4B2}

Next, we calculate the energy for the 2D SkL realized in Eq.~(\ref{eq:3Qansatz}) for  
$\Theta=\frac{\pi}{2}$. 
In this state, the energy depends on the relative phases between the constituent helices, 
as ${\bf Q}_\eta$ are not linearly independent. 
Thus, we need to extend Eq.~(\ref{eq:3Qansatz}) by replacing ${\bf Q}_\eta\cdot{\bf r}$ with 
${\bf Q}_\eta \cdot{\bf r} + \varphi_\eta$, where $\varphi_\eta$ denotes the phase shift for the ${\bf Q}_\eta$ component. 
For this state, there is an additional energy contribution from 
the quartic interaction $W\bigl({\bf S}({\bf r})\cdot{\bf S}({\bf r})\bigl)^2$, which is given by 
\begin{equation}
	f_W= 9Wm|\psi_{{\bf Q}_1}||\psi_{{\bf Q}_2}||\psi_{{\bf Q}_3}| \cos \sum_\eta \varphi_{\eta}.
	\label{eq:f_W}
\end{equation}
Thus, the energy of the SkL is given as 
\begin{equation}
F_{\rm SkL} = F_{\rm HL}+f_W.
\label{eq:F_SkL}
\end{equation}
Note that the relations in Eq.~(\ref{eq:HL_conditions}) hold also in the present case. 
We find that the energy is always optimized with $\sum_{\eta} \varphi_{\eta}=(2n+1)\pi$, where $n$ is an integer.

As for the HL above, the SkL turns into a topologically trivial state with $N_{\rm pair}=0$ for $m>\sqrt{3}$, 
while the energy can be estimated by Eq.~(\ref{eq:F_SkL}); see also the results in Sec.~\ref{sec:4.3}. 
We call such a topologically trivial state the 2D $3Q$ state in the following analysis.

\subsubsection{Helical or conical state}

Finally, we calculate the energy for the 1D helical or conical state in Eq.~(\ref{eq:3Qansatz}) with $\Theta=0$, 
where all three ${\bf Q}_\eta$ merge into one. 
The spin texture is given by 
\begin{equation}
	{\bf S}({\bf r}) = {\bf e}_1^1\cos{\bf Q}_1\cdot{\bf r}  
	+ {\bf e}_1^2\sin{\bf Q}_1\cdot{\bf r} + m\hat{{\bf z}}, 
\end{equation}
where ${\bf Q}_1 \parallel \hat{\bf z}$.
For this state, the energy is given by
\begin{equation}
F_{1Q} = 
f_2({\bf Q}_1) + f_s({\bf Q}_1) 
+ f_{m}^{(1)}(m)
+ f_{m}^{(2)}( {\bf Q}_{1}, m ),  
\label{eq:F_1Q}
\end{equation}
where $|\psi_{{\bf Q}_1}|=1$.

\subsection{Results \label{sec:4.3}}

\begin{figure}[t]
	\centering
	\includegraphics[width=1.0\columnwidth]{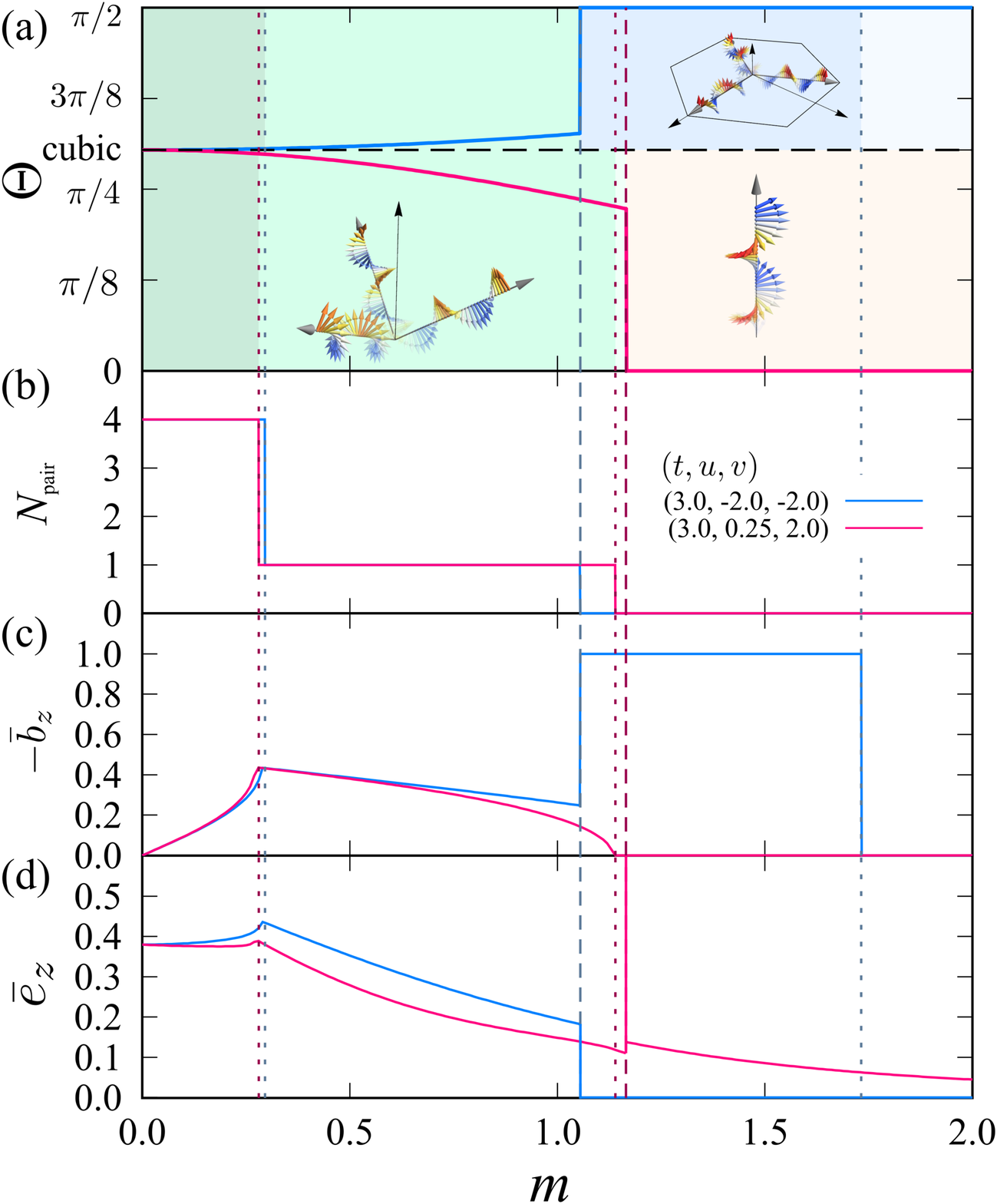}
	\caption{  \label{fig:vc}
	(a) Optimized value of $\Theta$ obtained by the variational calculations while changing $m$. 
	We show two cases starting from the 3D HL at $\Theta=\Theta_{\rm cubic}$ 
	for $m=0$: One with  $(t, u, v) = (3.0,-2.0,-2.0)$ exhibits a transition to $\Theta=\frac{\pi}{2}$ 
	(2D SkL with coplanar ${\bf Q}_\eta \perp \hat{\bf z}$), 
	and the other with $(t, u, v) = (3.0, 0.25 , 2.0)$ to $\Theta=0$ 
	(1D conical state with ${\bf Q}\parallel \hat{\bf z}$). 
	Schematic figures of each spin state are shown in the inset. 
	Corresponding values of (b) $N_{\rm pair}$,  (c) $-\bar{b}_z$, and (d) $\bar{e}_z$. 
	The green regions in (a) represent the states with $0<\Theta<\pi/2$; 
	the dark, normal, and pale colors are for the HLs with $N_{\rm pair}=4$ and $1$, 
	and the 3D $3Q$ state with $N_{\rm pair}=0$, respectively 
	[the 3D $3Q$ state is limited to the narrow region of $1.139\lesssim m\lesssim 1.164$ 
	for the case with $(t, u, v) = (3.0, 0.25 , 2.0)$]. 
	The dark and pale blue regions for the case with $(t, u, v) = (3.0, 0.25 , 2.0)$ represent 
	the SkL and the 2D $3Q$ states, respectively, at $\Theta=\pi/2$. 
	The red region for the case with $(t, u, v) = (3.0, 0.25 , 2.0)$ represents the $1Q$ conical state at $\Theta=0$. 
	The vertical dashed and dotted lines denote the magnetic and topological transitions, respectively. 
	}
\end{figure}

\begin{figure}[t]
\centering
\includegraphics[width=1.0\columnwidth]{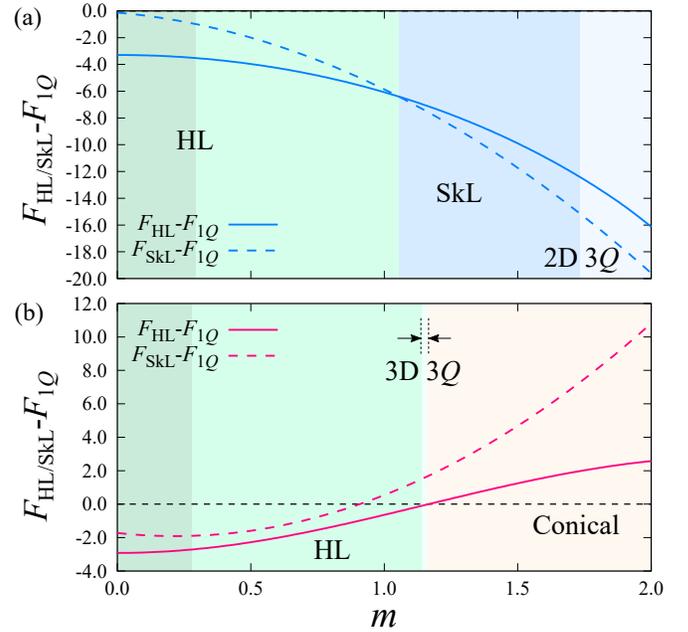}
\caption{
\label{fig:vc_ene}
Comparison of the energy between different states for (a) $(t,u,v) = (3.0,-2.0,-2.0)$ and (b) $(t,u,v)=(3.0,0.25,2.0)$. 
The solid (dashed) lines are the energy for the HL in Eq.~(\ref{eq:F_HL}) [SkL in Eq.~(\ref{eq:F_SkL})] measured from that for the $1Q$ conical state in Eq.~(\ref{eq:F_1Q}). 
The background colors are common to those in Fig.~\ref{fig:vc}(a). 
}
\end{figure}

Comparing the energies in Eqs.~(\ref{eq:F_HL}), (\ref{eq:F_SkL}), and (\ref{eq:F_1Q}), we determine the variational 
ground state.
We find a variety of modulations of the spin moir\'{e} depending on the parameters 
in the energy functional, $t$, $u$, and $v$ in Eq.~(\ref{eq:tuv}). 
Figure~\ref{fig:vc} shows two representative results starting from the cubic configuration with 
$\Theta=\Theta_{\rm cubic}$ at $m=0$. 
The parameters are taken as $(t,u,v)=(3.0,-2.0,-2.0)$, and $(t,u,v)=(3.0,0.25,2.0)$ for these two cases. 
The energy comparison corresponding to Fig.~\ref{fig:vc} is shown in Fig.~\ref{fig:vc_ene}.

For the case with $(t,u,v)=(3.0,-2.0,-2.0)$ denoted by the blue lines in Fig.~\ref{fig:vc}, 
$\Theta$ gradually increases with $m$ and changes discontinuously 
to $\Theta = \frac{\pi}{2}$ at $m\simeq 1.054$.
As the three ${\bf Q}_\eta$ become coplanar for $\Theta = \frac{\pi}{2}$, this is a magnetic phase transition 
to the 2D SkL; see the inset of Fig.~\ref{fig:vc}(a).
In this case, $N_{\rm pair}$ changes from $4$ to $1$ at $m\simeq 0.295$, and from $1$ to $0$ 
at the magnetic transition to the SkL, as shown in Fig.~\ref{fig:vc}(b). 
The former is a topological transition within the HL phase, where $\bar{b}_z$ shows a cusp. 
In the magnetic transition at $m\simeq 1.054$, $\bar{b}_z$ jumps to a quantized value $-1$ in the SkL state, 
while it vanishes for $m > \sqrt{3}$, as shown in Fig.~\ref{fig:vc}(c); 
the system undergoes another transition to the 2D $3Q$ state at $m=\sqrt{3}$, where the $z$ component of the core spins are turned 
from negative to positive and the spin texture becomes topologically trivial. 
Meanwhile, $\bar{e}_z$ also shows a cusp at the topological transition, but it vanishes at the magnetic transition 
to the SkL with showing a negative $\delta$-function anomaly, as shown in Fig.~\ref{fig:vc}(d). 
Note that $\bar{e}_z$ includes both ${e}_z^m({\bf r})$ and ${e}_z^\Theta({\bf r})$ in Eq.~(\ref{eq:ezave2}), as $\Theta$ changes as $m$, 
in contrast to the calculations in Sec.~\ref{sec:3.2}. 
This anomaly comes from the discontinuous change of $\Theta$.

On the other hand, for the case with $(t,u,v)=(3.0,0.25,2.0)$ denoted by the red lines in Fig.~\ref{fig:vc}, 
$\Theta$ decreases gradually until the sudden drop to $0$ at $m\simeq 1.164$. 
As all ${\bf Q}_\eta$ overlap for $\Theta=0$, this is a magnetic phase transition to 
a 1D conical state specified by the single ${\bf Q}\parallel\hat{{\bf z}}$; see the inset of Fig.~\ref{fig:vc}(a).
In this case, $N_{\rm pair}$ changes from $4$ to $1$ at $m\simeq 0.280$ and from $1$ to $0$ at $m\simeq 1.139$; 
namely, the system undergoes two successive topological transitions before the magnetic transition 
to the conical state. 
The narrow window for $1.139\lesssim m\lesssim 1.164$ is a topologically-trivial 3D $3Q$ state with $N_{\rm pair}=0$. 
This is in contrast to the above case with $(t,u,v)=(3.0,-2.0,-2.0)$ where the topological transition with $N_{\rm pair}=1\to 0$ 
occurs simultaneously with the magnetic transition to the SkL. 
In addition to the cusp at the topological transition at $m\simeq 0.280$, 
$\bar{e}_z$ shows a small hump at $m\simeq 1.1$ before the second topological transition, 
which is ascribed to the contribution from $e_z^{\Theta}({\bf r})$.  
$\bar{e}_z$ remains nonzero in the conical state after showing a positive $\delta$-function anomaly 
at the magnetic phase transition.

\section{Discussion \label{sec:5}}

Through this study, we clarified the systematic evolution of the net values of the emergent electromagnetic fields, $\bar{\bf b}$ and $\bar{\bf e}$, while changing the spin moir\'e, by establishing their relations to the topological objects, i.e., the hedgehogs, antihedgehogs, and Dirac strings. 
This would shed light on the experimental detection of the topological properties of the system through 
electromagnetic responses. 
For instance, as $\bar{\bf b}$ manifests in the topological Hall effect, the results in Fig.~\ref{fig:vc} 
indicates that the cusp in the topological Hall conductivity would be an indicator of the topological transition 
by pair annihilation of monopoles and antimonopoles in the HL phase. 
Meanwhile, $\bar{\bf e}$, which is relevant to dynamical electric responses to time-dependent modulation of the spin moir\'{e}, 
would also be useful to detect the topological transitions as indicated by our results.
We note that an inductance generated by the emergent electric field was recently discussed for current-driven dynamics in spin helices~\cite{Nagaosa2019,Yokouchi2020}. 
Our results indicate the possibility that such dynamical phenomena can be driven by not only an electric current 
but also a magnetic field in the HL.
Thus, our findings would push forward the exploration of the emergent electromagnetic phenomena in a wider variety of topological spin textures. 

In the present study, we dealt with the spin moir\'{e} with two parameters $\Theta$ and $m$, but one can extend the analysis 
to other parameters, such as the periods, amplitudes, phases, and the number of constituent helices by using the analogy with the conventional moir\'e. 
We note that the phase degree of freedom was recently studied for 2D SkLs~\cite{Hayami2020phase} and 
the amplitude differences were also studied for superpositions of three and four helices in three dimensions~\cite{Shimizu2021anisotropy}.
Also, as mentioned in the end of Sec.~\ref{sec:2}, the present analysis includes the effect of the tilting of the helical planes partly. 
Given many examples in optics, these parameters give rise to further variety of magnetic textures and quantum phases. 
This approach based on the spin moir\'e picture will be efficient and useful for designing the emergent electromagnetism and 
the resultant electronic, transport, and optical properties.

Experimentally, the spin moir\'{e} could be modulated, e.g., by an external magnetic field, pressure, and chemical substitution. 
Interestingly, a modulation of $\Theta$ was observed by changing the sample thickness in MnGe~\cite{Kanazawa2017}. 
Furthermore, a recent experiment for MnGe showed that $\Theta$ decreases as increasing a magnetic field~\cite{Kanazawa2020}. 
Our results would provide a starting point for understanding these findings.  
It would also be intriguing to discuss the topological Hall effect~\cite{Kanazawa2011,Kanazawa2016, Kanazawa2020, Fujishiro2020AHE} 
and the large thermoelectric effect~\cite{Shiomi2013,Fujishiro2018} in the magnetic field from the viewpoint of the spin moir\'{e} 
proposed here. 

\section{Summary \label{sec:6}}

To summarize, we have theoretically studied the effect of the angles between the helical directions and the net magnetization 
on the spin moir\'e composed of three spin helices.  
We found topological transitions between the 3D HLs with different numbers of the magnetic monopole-antimonopole pairs 
and the topologically trivial states, in addition to the magnetic transitions to a 2D SkL and a 1D conical state. 
By tracking the monopoles and antimonopoles, the Dirac strings, and the emergent electromagnetic fields 
in real space, we clarified the details of the topological transitions caused by pair annihilation of monopoles and antimonopoles. 
We also calculated how the net values of the emergent magnetic and electric fields change through these transitions, 
which are important to the bulk properties of the system. 
We found that both values are maximized with showing a cusp at the topological transition with annihilations of 
three out of four monopole-antimonopole pairs while increasing the magnetization. 
In addition, we also performed a variational calculation based on the Ginzburg-Landau free energy to demonstrate the phase transitions expected for an applied magnetic field. 
Starting from the cubic HL with zero magnetization, we discussed the two representative cases: 
One is the case in which the angle spanning the helical directions increases with the magnetization and the system turns into 
the 2D SkL, and the other is that the angle decreases and the system turns into the 1D conical state. 
In both cases, the HL state experiences topological transitions by pair annihilation of monopoles and antimonopoles, where the emergent magnetic and electric fields show characteristic anomalies. 
Our results lay the basis for efficient engineering of the emergent electromagnetism and topological nature 
through the manipulation of the spin moir\'{e}.

In the present study, we assumed the phenomenological energy functional following the previous studies in Refs.~\cite{Binz2006-1,Binz2006-2}. 
It is crucially important to study a microscopic model beyond the phenomenological arguments, 
especially for understanding the recent experimental results described in Sec.~\ref{sec:5}.
Generally speaking, multiple-spin interactions, which are relevant to 
$F_4$ in Eq.~(\ref{eq:F4}), play an important role in the stabilization of noncoplaner spin textures~\cite{Akagi2012, Muhlbauer2009, Binz2006-1, Binz2006-2, Park2011, Hayami2017, Grytsiuk2020, Okumura2020}.
It is desired to identify the relevant microscopic interactions for the spin moir\'e engineering in actual materials. 
Another interesting issue is to explore new phenomena related with the emergent electric field from the spin moir\'e modulation. 
Our results suggest that a time-dependent external magnetic field can generate the emergent electric field in the 3D 
spin moir\'es through the movement and pair annihilation of monopoles and antimonopoles. 
Further theoretical and experimental studies are desirable for clarifying such a new type of dynamical magnetoelectric effects.

\begin{acknowledgments}
The authors thank S. Hayami, S.-H. Jang, N. Kanazawa, J. Masell, and K. Nakazawa for fruitful discussions. 
This research was supported by Grant-in-Aid for Scientific Research Grants Numbers 
JP18K03447, JP19H05822 and JP19H05825, and JST CREST (JP-MJCR18T2), and 
the Chirality Research Center in Hiroshima University and 
JSPS Core-to-Core Program, Advanced Research Networks.
K.S. was supported by the Program for Leading Graduate Schools (MERIT-WINGS). 
S.O. was supported by JSPS through the research fellowship for young scientists. 
Y.M. acknowledges stimulated discussions in the meetings of the
Cooperative Research Project of the Research Institute of Electrical
Communication, Tohoku University. 
Parts of the numerical calculations were performed in the supercomputing systems in ISSP, the University of Tokyo.
\end{acknowledgments}

\appendix*

\section{Tilting of the helical planes}

In this Appendix, we show that the effect of a tilting of the helical planes in Eq.~(\ref{eq:3Qansatz}) is understood by a simple scaling of the results for Eq.~(\ref{eq:3Qansatz}) obtained in the main text, 
as long as the threefold rotational symmetry about the $z$ axis is preserved.
Such a tilting is decomposed into two types: 
One is simultaneous rotation of the three normal vectors ${\bf e}_{\eta}^0$ about the $z$ axis, and the other is simultaneous tilting of ${\bf e}_{\eta}^0$ along the $\Theta$ direction. 
The former is trivial as it is reduced to global spin rotation about the $z$ axis. 
To discuss the effect of the latter, we define the titling angle $\Phi$ to satisfy $\hat{\bf z}\cdot\tilde{{\bf e}}_{\eta}^0=\cos(\Theta-\Phi)$, where $\tilde{\bf e}_{\eta}^0$ are the normal vectors of the tilted helical planes ($0\leq \Phi \leq \pi$).
Then, the spin structure obtained by the superposition of the tilted helices is represented by 
\begin{eqnarray}
{\bf S}({\bf r}) =
\sum_{\eta=1}^{3} \frac{1}{\sqrt{3}}\left(
\tilde{\bf e}_{\eta}^{1}\cos\QQ_{\eta}
+\tilde{\bf e}_{\eta}^{2}\sin\QQ_{\eta}
\right)+m\hat{{\bf z}}, \label{eq:app_3Qansatz}
\end{eqnarray}
where $\tilde{\bf e}_{\eta}^{1}$ are obtained by replacing $\Theta$ with $\Theta-\Phi$ in Eqs.~(\ref{eq:e_1^1}), (\ref{eq:e_2^1}), and (\ref{eq:e_3^1}), and $\tilde{\bf e}_{\eta}^{2}$ satisfy $\tilde{\bf e}_{\eta}^{1}\times\tilde{\bf e}_{\eta}^{2}=\tilde{\bf e}_{\eta}^{0}$; the other notations are common to those in Eq.~(\ref{eq:3Qansatz}). 

We show that the tilted spin structure in Eq.~(\ref{eq:app_3Qansatz}) can be rescaled to a superposition of proper screws. 
This is done by a simple scaling of the real-space coordinate as 
$(\tilde{x},\tilde{y},\tilde{z}) = (R_s x, R_s y, R_c z)$, where 
\begin{align}
R_s = \frac{\sin\Theta}{\sin(\Theta-\Phi)}, \ \  
R_c = \frac{\cos\Theta}{\cos(\Theta-\Phi)}.
\end{align} 
In this new coordinate, the spin structure in Eq.~(\ref{eq:app_3Qansatz}) is described as 
\begin{eqnarray}
\tilde{\bf S}(\tilde{\bf r}) =
\sum_{\eta=1}^{3} \frac{1}{\sqrt{3}}\left(
\tilde{\bf e}_{\eta}^{1}\cos\tilde{\QQ}_{\eta}
+\tilde{\bf e}_{\eta}^{2}\sin\tilde{\QQ}_{\eta}
\right)+m\hat{{\bf z}}, \label{eq:app_3Qansatz_map}
\end{eqnarray}
where $\tilde{\QQ}_{\eta} = \tilde{\bf Q}_{\eta}\cdot\tilde{\bf r}$,  
and $\tilde{\bf Q} = Q\tilde{\bf e}_{\eta}^0$, and $\Theta-\Phi \neq \frac{n}{2}\pi$ ($n$ is an integer).
This indicates that $\tilde{\bf S}(\tilde{\bf r})$ is composed of the proper screws, similar to ${\bf S}({\bf r})$ in Eq.~(\ref{eq:3Qansatz}).
Hence, the magnetic and topological properties of Eq.~(\ref{eq:app_3Qansatz}) are obtained from those of 
Eq.~(\ref{eq:app_3Qansatz_map}), which are calculated for Eq.~(\ref{eq:3Qansatz}) in the main text. 
For instance, the emergent electromagnetic fields for Eq.~(\ref{eq:app_3Qansatz}) are obtained by those for Eq.~(\ref{eq:app_3Qansatz_map}) as 
\begin{align}
&{\bf b}({\bf r}) = ( 
R_s R_c \tilde{b}_{\tilde{x}}(\tilde{\bf r}), 
R_s R_c \tilde{b}_{\tilde{y}}(\tilde{\bf r}), 
R_s^2 \tilde{b}_{\tilde{z}}(\tilde{\bf r}) 
), \\
&{\bf e}^\nu({\bf r}) = (
R_s \tilde{e}_{\tilde{x}}^\nu(\tilde{\bf r}), 
R_s \tilde{e}_{\tilde{y}}^\nu(\tilde{\bf r}), 
R_c \tilde{e}_{\tilde{z}}^\nu(\tilde{\bf r}) 
), 
\end{align} 
where 
$\tilde{\bf b}(\tilde{\bf r})$ and $\tilde{\bf e}^{\nu}(\tilde{\bf r})$ are defined by replacing ${\bf n}({\bf r})$ with 
$\tilde{\bf n}(\tilde{\bf r})=\tilde{\bf S}(\tilde{\bf r})/|\tilde{\bf S}(\tilde{\bf r})|$ and taking the partial derivatives 
in terms of $\tilde{\bf r}$ in Eqs.~(\ref{eq:b_cont}) and (\ref{eq:ezave}). 
We note that the averages $\bar{b}_z$ and $\bar{e}_z$ for Eq.~(\ref{eq:app_3Qansatz}) are equivalent to the integrals of $\tilde{\bf b}(\tilde{\bf r})$ and $\tilde{\bf e}^\nu(\tilde{\bf r})$ within the MUC in the new coordinate.

\bibliography{spinmoire}

\end{document}